\documentstyle{mn}

\input{epsf}

\newif\ifAMStwofonts

\def\LCDM{$\Lambda$CDM}
\def\apj{ApJ}
\def\mnras{MNRAS}
\def\apjl{ApJL}
\def\aap{A\&A}

\def\Mpc{\ {\rm Mpc}}

\newcommand{\be}{\begin{equation}}
\newcommand{\eq}{\end{equation}}

\newcommand{\Msun}{{\rm M_\odot}}

\ifoldfss
  \ifCUPmtlplainloaded \else
    \NewTextAlphabet{textbfit} {cmbxti10} {}
    \NewTextAlphabet{textbfss} {cmssbx10} {}
    \NewMathAlphabet{mathbfit} {cmbxti10} {} 
    \NewMathAlphabet{mathbfss} {cmssbx10} {} 
  \fi
  \ifAMStwofonts
    \ifCUPmtlplainloaded \else
      \NewSymbolFont{upmath} {eurm10}
      \NewSymbolFont{AMSa} {msam10}
      \NewMathSymbol{\upi}     {0}{upmath}{19}
      \NewMathSymbol{\umu}     {0}{upmath}{16}
      \NewMathSymbol{\upartial}{0}{upmath}{40}
      \NewMathSymbol{\leqslant}{3}{AMSa}{36}
      \NewMathSymbol{\geqslant}{3}{AMSa}{3E}

      \let\leq=\leqslant 
       
    \fi
  \fi
\fi 

\ifnfssone
  \newmathalphabet{\mathit}
  \addtoversion{normal}{\mathit}{cmr}{m}{it}
  \addtoversion{bold}{\mathit}{cmr}{bx}{it}
  \newmathalphabet{\mathbfit} 
  \addtoversion{normal}{\mathbfit}{cmr}{bx}{it}
  \addtoversion{bold}{\mathbfit}{cmr}{bx}{it}
  \newmathalphabet{\mathbfss} 
  \addtoversion{normal}{\mathbfss}{cmss}{bx}{n}
  \addtoversion{bold}{\mathbfss}{cmss}{bx}{n}
  \ifAMStwofonts
    \ifCUPmtlplainloaded \else
      \UseAMStwoboldmath
      \makeatletter
      \new@mathgroup\upmath@group
      \define@mathgroup\mv@normal\upmath@group{eur}{m}{n}
      \define@mathgroup\mv@bold\upmath@group{eur}{b}{n}
      \edef\UPM{\hexnumber\upmath@group}
      \new@mathgroup\amsa@group
      \define@mathgroup\mv@normal\amsa@group{msa}{m}{n}
      \define@mathgroup\mv@bold\amsa@group{msa}{m}{n}
      \edef\AMSa{\hexnumber\amsa@group}
      \makeatother
      \mathchardef\upi="0\UPM19
      \mathchardef\umu="0\UPM16
      \mathchardef\upartial="0\UPM40
      \mathchardef\leqslant="3\AMSa36
      \mathchardef\geqslant="3\AMSa3E

      \let\leq=\leqslant 

    \fi
  \fi
\fi 

\ifnfsstwo
  \DeclareMathAlphabet{\mathbfit}{OT1}{cmr}{bx}{it}
  \SetMathAlphabet\mathbfit{bold}{OT1}{cmr}{bx}{it}
  \DeclareMathAlphabet{\mathbfss}{OT1}{cmss}{bx}{n}
  \SetMathAlphabet\mathbfss{bold}{OT1}{cmss}{bx}{n}
  \ifAMStwofonts
    \ifCUPmtlplainloaded \else
      \DeclareSymbolFont{UPM}{U}{eur}{m}{n}
      \SetSymbolFont{UPM}{bold}{U}{eur}{b}{n}
      \DeclareSymbolFont{AMSa}{U}{msa}{m}{n}
      \DeclareMathSymbol{\upi}{0}{UPM}{"19}
      \DeclareMathSymbol{\umu}{0}{UPM}{"16}
      \DeclareMathSymbol{\upartial}{0}{UPM}{"40}
      \DeclareMathSymbol{\leqslant}{3}{AMSa}{"36}
      \DeclareMathSymbol{\geqslant}{3}{AMSa}{"3E}

      \let\leq=\leqslant 

    \fi
  \fi
\fi 

\ifCUPmtlplainloaded \else
  \ifAMStwofonts \else 
    \def\upi{\pi}
    \def\umu{\mu}
    \def\upartial{\partial}
  \fi
\fi

\title[Galaxy satellites in \LCDM~hydrodynamical simulations]{Radial distribution and strong lensing statistics of satellite galaxies
  and substructure using high resolution \LCDM~hydrodynamical simulations}

\author[A.V. Macci\`o, B. Moore, J. Stadel $\&$ J. Diemand]
{Andrea V. Macci\`o$^1$ \thanks{andrea@physik.unizh.ch},
Ben Moore$^1$, Joachim Stadel$^1$  \&  J\"urg Diemand$^{1,2}$\\
$^1$Institute for Theoretical Physics, University of Z\"urich,
Winterthurerstrasse 190 ,CH-8057 Z\"urich, Switzerland \\
$^2$Department of Astronomy \& Astrophysics, University of California, 1150 High
Street, Santa Cruz CA 95064, USA
}

\smallskip 

\pubyear{2005}

\begin{document}

\maketitle


\begin{abstract}

We analyse the number density and radial distribution of substructures and satellite galaxies
using cosmological simulations that follow the gas dynamics of 
a baryonic component, including shock heating, radiative
cooling and star formation within the hierarchical concordance \LCDM~ model. 
We find that the dissipation of the baryons greatly enhances the survival
of subhaloes, expecially in the galaxy core, resulting in a radial
distribution of satellite galaxies that closely follows the overall mass
distribution in the inner part of the halo.
Hydrodynamical simulations are necessary to resolve the adiabatic contraction
and dense cores of galaxies, resulting in a total number of satellites a factor of two
larger than found in pure dark matter simulation, in good agreement with the
observed spatial distribution of satellite galaxies within galaxies and clusters. 
Convergence tests show that the cored distribution found by previous authors
in pure N-body simulations was due to physical overmerging of dark matter only structures.

We proceed to use a ray-shooting technique in order to study the impact of these
additional substructures on the number of violations of the cusp caustic magnification
relation. We develop a new approach to try to disentangle the effect of 
substructures from the intrinsic discreteness of N-Body simulations.
Even with the increased number of substructures in the centres
of galaxies, we are not able to reproduce the observed high numbers of discrepancies observed
in the flux ratios of multiply lensed quasars.

\end{abstract}


\section{INTRODUCTION}
\label{section:introduction}

The study of substructures within simulated galaxy sized haloes have posed interesting
problems to the current, widely accepted, \LCDM~ scenario.
As pointed out previously, (Klypin et al 1999, Moore et al. 1999) the number
of surviving subhaloes
found in N-Body simulations greatly exceeds the number of observed satellites 
around the Milky Way and Andromeda. The properties of subhaloes on different scales
has been the subject of many recent studies which have pushed the resolution of
dissipationless simulations (Moore et al. 1998, 
Ghigna et al. 1998, Colin et al. 2000, Ghigna et
al 2000, Springel et al. 2001, De Lucia et al. 2004, Kravtsov et al 2004, 
Diemand et al 2004a, Gao et al 2004a, Reed et al 2005, Zentner et al 2005a,
Zentner et al 2005b). 
The kinematical properties of subhaloes are now well understood -
they make up a fraction of between 5 and 10\% of the mass of virialized haloes
on scales relevant to observational cosmology.

Most of these previous studies used dissipationless cosmological simulations;
although non-baryonic dark matter exceeds baryonic matter by a factor of
$\Omega_{dm}/\Omega_b \simeq 6$ on average, the gravitational field in the
central region of galaxies is dominated by stars and gas. In the hierarchical galaxy
formation model, stars are formed by the condensation of cooling
baryons at the halo center. The cooling baryons increase the density in 
the central halo region mainly because of the extra mass associated with the inflow,
but also because of the adiabatic contraction of the total mass distribution
(Blumenthal et al (1986), Loeb \& Peebles (2003), Gao et al. (2004b), Gnedin et al. (2004)).
This process acts {\it both} for the host halo and its subhaloes, therefore we
may expect that the dark matter substructure haloes formed within 
hydrodynamical simulations will experience a different
tidal force field and they themselves will be more robust to tidal effects.
Since overmerging of dark matter substructures is sensitive to their central structure
(Moore, Katz \& Lake 1996), pure N-Body simulations may bias results because of physical
overmerging (Diemand, Moore \& Stadel 2004a, DMS04 hereafter).

The fact that substructure haloes are spatially more extended than the averaged
mass distribution was first pointed out by Ghigna et al. (1998). By increasing the
resolution by over an order of magnitude, DMS04 showed that 
this result was independent of the numerical resolution. Furthermore the 
distribution of galaxies in clusters appears to follow the overall mass distribution, quite
unlike the subhaloes selected by final bound mass in
pure dark matter simulations. By tracing haloes backwards and
forwards in time through the simulations they argued that the missing central subhaloes may be
destroyed by tidal forces at redshifts higher than z=5.
The survival of substructure and galaxies within dense environments has implications for 
indirect detection techniques, for example using image distortions of
gravitationally lensed distant quasars by foreground galaxies and their dark
matter haloes.

It has been argued that a possible signature of the presence of dark matter
substructures can be found in strong gravitational lensing of QSOs 
(Mao \& Schneider 1998; Metcalf \& Madau 2001; Chiba 2002; 
Metcalf \& Zhao 2002; Dalal \& Kochanek 2004; Chen, Kravtsov \& Keeton 2003; 
Kochanek \& Dalal 2004; Mao et al 2004, Amara 2004). 
If a distant image source is close to (or inside) a cusp in a caustic
curve, three of the images will be clustered together and the sum of their
magnifications will be zero (Zakharov (1995), taking the negative parity image to have
negative magnification). This relation holds for a wide class of smooth
analytic lens models (i.e. Schneider \& Weiss 1992, Keeton et al. 2003); 
on the other hand all known observed
lensed QSOs violate this relation which has been explained as due to the presence of
cold dark matter substructure within the lensing galaxy's halo.
However, some of these discrepant systems may be due to
microlensed stars rather than to cold dark matter substructure (Keeton et al
2003). Bradac et al. (2004) analysed a low resolution simulation of a 
galaxy and claim that the level of substructure present in simulations produces violations of
the cusp relation comparable to those observed. 
Amara et al. (2004) implanted an idealised model of galaxy into the center
of a high resolution galactic halo extracted from dissipationless N-Body
simulations to test the effects of substructure on lensed images.
Their findings contrast those of Bradac et al 2004,
since they found that the substructures produced in a \LCDM~ halo are not abundant
enough to account for the cusp caustic violation observed, these results are
also confirmed by Mao et al 2004, based on analysis of substructure abundance
in pure dark matter simulations.

The first part of this paper is devoted to the analysis of the subhalo
population around a galaxy sized halo that forms in a large cosmological simulation
simulated with dark matter only and then with the inclusion of a baryonic component.
In section \ref{sec:num_sim} we present the numerical simulations, our 
halo--finding scheme and resolution tests. The properties of the main halo and its
satellites are presented in section \ref{sec:HalPro}.
Comparison with observations and a discussion of the results are in section \ref{sec:Obs}.
In the second part of this paper we re-examine the effects of substructures
on multiply lensed quasar images using our new
hydrodynamic galaxy simulation in combination with a ray shooting
technique. In section \ref{sec:Lenscode} we present the lensing code,
while section \ref{sec:cusp} is devoted to multiple images analysis and in sections
\ref{sec:LensRes} and \ref{sec:Concl}
we present our results and the conclusions of our work.

\section{NUMERICAL SIMULATIONS}
\label{sec:num_sim}

The simulations were performed with GASOLINE, a multi-stepping, parallel 
TreeSPH $N$-body code (Stadel 2001, Wadsley et al. 2004).
We include radiative and Compton cooling for a primordial 
mixture of hydrogen and helium. The star formation algorithm is based on 
a Jeans instability criteria (Katz 1992), where gas particles in dense, 
unstable regions and in 
convergent flows spawn star particles at a rate proportional to the local dynamical
time (see also Governato et al 2004). The star formation efficiency
was set to $0.1$, but in the adopted scheme its precise value has only a minor effect 
on the star formation rate (Katz 1992). The code also includes supernova
feedback as described by (Katz 1992), and a UV background following Haardt \&
Madau 1996.

We have selected a candidate Galactic mass halo ($M_{dm} \approx 10^{12} \Msun$) from an
existing low resolution dark
matter only simulation in a concordance ($\Lambda$=0.7,$\Omega_0$=0.3,
$\sigma_8$=0.9) cosmology and resimulated it at higher resolution using the
volume renormalization  technique (Katz \& White 1993), and including a gaseous component
within the entire high resolution region. The mass per particle of the dark matter
and gaseous particles are respectively
$m_{d} = 1.66 \times 10^6 M_{\odot}$ and $m_g = 3.28 \times 10^5
M_{\odot}$. The dark matter has a spline gravitational softening length of 200 pc and we
have about $1\times 10^6$ particles for each component (dark and gas) in the
high resolution region.

The SPH simulations with cooling are computationally expensive and we are 
forced to stop the full calculation once the galaxy has formed, at a redshift 
$z=1.5$. (The parallel calculation is dominated by the few remaining gas particles which
need extremely small timesteps in order to satisfy the Courant criterion). 
In order to study the dynamical evolution of the stellar 
and dark matter satellites we have evolved the simulation to the present epoch
without following the remaining gaseous particles - we turn them into collisionless
particles and treat only their gravitational interactions. We do not believe that
this influences any of our conclusions since (i) most of the gas has already  turned into
stars between a redshift z=5.5 and z=2.5, 
and (ii) continuing to include cooling of the remaining gas would only
allow us to resolve higher density gas clouds. 
Inside the virial radius at a redshift z=0 we have $8.6 \times 10^5$ dark
matter particles and  $2.4 \times 10^5$ star particles.
In Figure \ref{fig:SFR} we show the star
formation rate in our simulation which is in good agreement with previous numerical 
studies (i.e. Governato et al. 2004).

The same object has also been simulated with dark matter only using the same
spatial and mass resolution as the hydro run, in addition one simulation with
four times better mass resolution (this object is Gal1 in DMS04).
In the following we will refer to the three simulations as {\it hydro}, {\it dm}
and $dm_{HR}$ respectively.

\section{Haloes properties}
\label{sec:HalPro}

\subsection{Host halo}

\begin{figure}
\centering
\epsfxsize=\hsize\epsffile{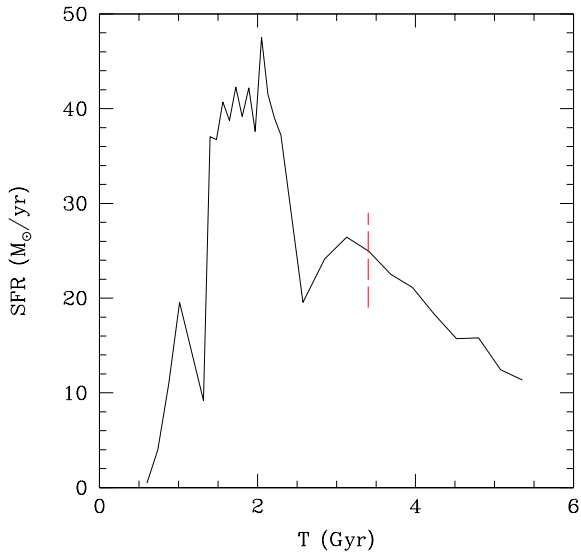}
\caption{
The star formation rate in the hydro simulation. 
The time since the big-bang is indicated and the red vertical line corresponds
to $z=2$, where cooling were switched off.}
\label{fig:SFR}
\end{figure}

\begin{figure}
\centering
\epsfxsize=\hsize\epsffile{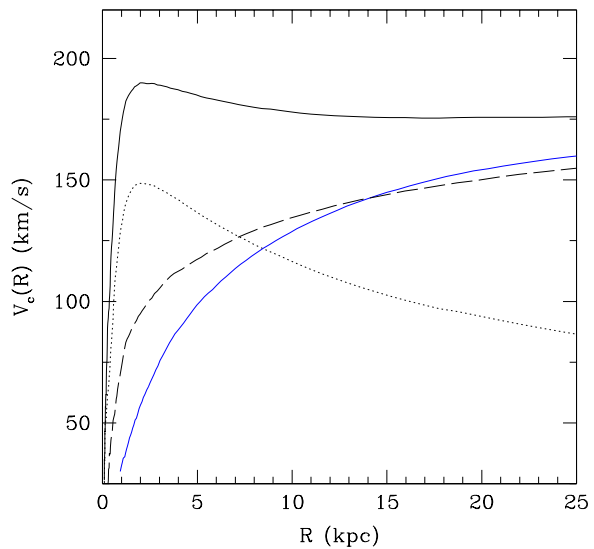}
\caption{Rotation curve (defined as $V_c(R) = \sqrt{GM(<R)/R}$) for all
  components (upper solid line), star (dotted line) and dark matter (dashed
  line) for the galaxy at z=0. For comparison the dark matter
  rotation curve (lower solid line) obtained in the pure dark matter
  simulation of the same object is also shown.}
\label{RotCurv}
\end{figure}

\begin{figure}
\centering
\epsfxsize=\hsize\epsffile{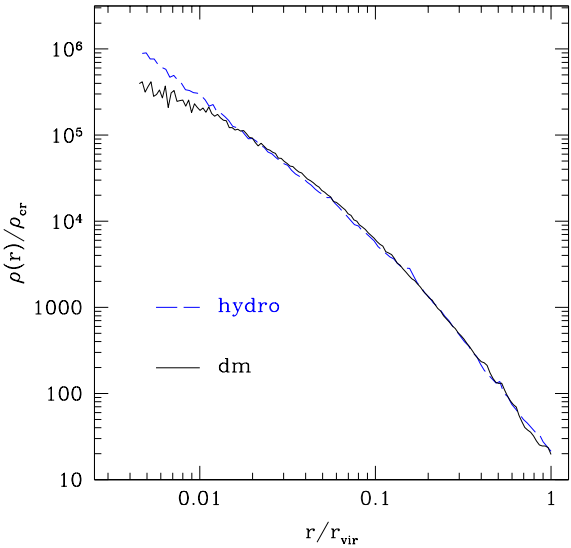}
\caption{Dark matter density profile in the N-Body (dashed curve) and Hydro
  (solid line) simulations, the effect of the adiabatic contraction can be
  clearly seen in the inner part of the profile.}
\label{fig:dprof}
\end{figure}

In Figure \ref{RotCurv} we show the rotation curve 
(defined as $V_c(r) = \sqrt{GM(<r)/r}$) for all components, stars
and dark matter separately for the galaxy at z=0. For comparison we also show the
dark matter rotation curve obtained in the pure dark matter simulation of the same object.
The baryons dominate within the inner 10 kpc and the effect of the adiabatic
contraction on the dark matter can be clearly seen - the baryons have increased
the mass within 5 kpc by a factor of four.
The steeper central cusp can be clearly seen in the density profile
plot (fig. \ref{fig:dprof}), the {\it dm} halo has a profile that can be well
fitted by an NFW profile with a concentration $c_{vir} = 9.6$ 
(defined with respect to the virial radius); on the other hand in the presence of gas
and stars the density profile in the inner region has an almost constant
cusp slope $\alpha \approx  -2.0$, as predicted by the adiabatic contraction
model (Blumenthal et al. 1986).
In figure \ref{fig:SFR} we show the star formation rate (SFR), the
bulge of the galaxy forms through a series of rapid major merger events that end
around $z=2.8$, turning mostly low angular momentum gas into stars.
There are no more merging events in the formation of our galaxy after $z=2$
so even if we cannot follow the SFR directly beyond $z=1.5$ we do not expect it to be
different from a slowly decreasing function of time.


\subsection{Sub haloes}
\label{sec:sub}

Within the virial radius of the high resolution CDM simulations
we can resolve several hundreds of substructure haloes (bound over-dense 
clusters of particles).
We identify subhaloes with SKID (Stadel 2001), which calculates local densities
using an SPH kernel, then moves particles along the density gradient
until they oscillate around a point (i.e. move less than some length $l$).
Then they are linked together using FoF with this $l$ as a linking length. 

SKID with $l = 4 \epsilon_0$ (where $\epsilon_0$ is the gravitational 
softening of the simulation) adequately identifies the smallest subhaloes 
and the centers of the largest subhaloes. For the latter the calculated bound mass is
underestimated. Using $l = 10 \epsilon_0$ can cure this, but then some of
the small subhaloes are missed. Therefore we have used a combination of the subhalo
catalogues obtained with these two linking lengths in order to create the
complete catalogue of subhaloes and to calculate their correct structural parameters.
We included in our study only halos with $M > 2 \times 10^8 \Msun (i.e. n_{dm} > 100)$.

\begin{figure}
\centering
\epsfxsize=\hsize\epsffile{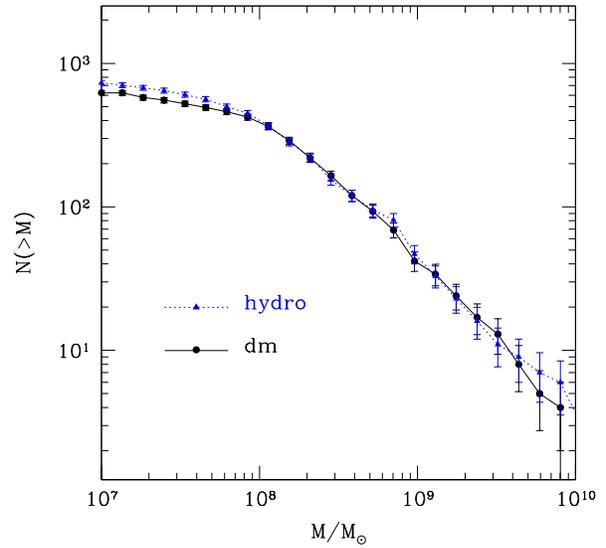}
\caption{Dark matter subhaloes mass function in the {\it hydro} (black) and
  {\it dm} (blue) simulations, in a sphere of $1$ \Mpc~ around the center of the galaxy.}
\label{fig:mfsub}
\end{figure}

\begin{figure}
\centering
\epsfxsize=\hsize\epsffile{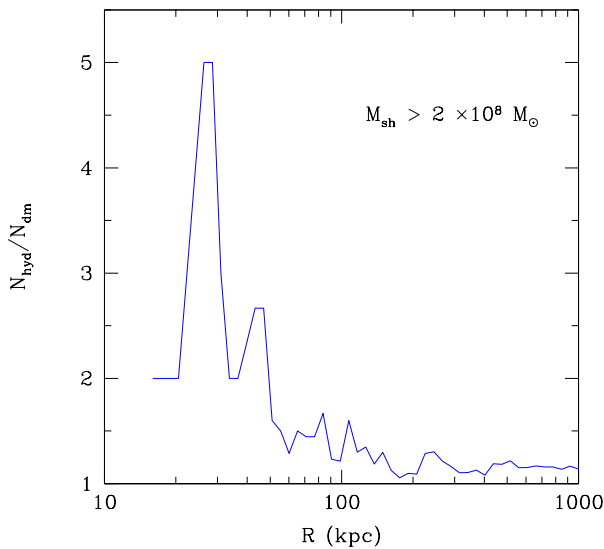 }
\caption{Ratio between the number of DM subhaloes with $M>2 \times 10^8 \Msun$ in the
  {\it hydro} and {\it dm} runs as function of the distance from the center of the main halo.}
\label{fig:mfR}
\end{figure}

In Figure \ref{fig:mfsub}, where we plot the dark
matter subhaloes mass function in the {\it hydro} and {\it dm} simulations, calculated
within a sphere of radius 1 Mpc centred on the galaxy. As expected, we do not
observed a big difference between the two simulations in such a large
volume, this is because the presence of baryons is thought to be much more important
in the inner region of our galaxy where the tidal field is much more efficient
in destroying subhaloes. 
The effect of baryons can be easily seen in Figure \ref{fig:mfR}, where we
plot the ratio of the number of subhaloes in the two simulations as function of
the distance from
the center of the galaxy. At 70 kpc ($\approx 1/3R_{200}$) we see an increase by a factor
of two in the numbers of surviving satellites. Within the inner 40 kpc the number of
satellites with mass 
greater than $2 \times 10^8 \Msun$ is enhanced by a factor 4-5 over that found in the
pure dark matter simulation.
We found that this difference is more striking for the most massive subhaloes,
the number of satellites with mass greater than $10^7 \Msun$ does not differ
so much from the N-body result.
This may be revealing the limitation
of our resolution - the smallest haloes of mass $\sim 10^7M_\odot$ are just resolved
with about 10 DM particles and are more easily destroyed through numerical 
effects. 
Another explanation could be the inhibition of star formation in smaller haloes
because of the ultra-violet (UV) background and supernovae (SN) feedback that raise the
temperature of the gas and stop its collapse in the DM halo if it is not
sufficiently bound (i.e. its potential well is not deep enough). 
Without stars, the DM can be more easily destroyed by the
tidal forces of the main halo (see also section \ref{sec:Obs} for more
details on the feedback effects).

As noticed by previous authors (DMS04 for a recent analysis) the spatial
distribution of subhaloes in cold dark matter simulations of galaxies is
anti-biased with respect to the mass; Nagai \& Kravtsov (2005) have 
recently shown that part of this bias is due to the varying amount of mass
loss at different radii, and that it is considerably smaller if instead of
using the mass of satellites at $z=0$, one uses the mass measured at the
accretion time. For our purposes we have decide to use the mass at the present
time, because we are interested in the different tidal forces and mass loss 
between the {\it hydro} and {\it dm} run.

\begin{figure}
\centering
\epsfxsize=\hsize\epsffile{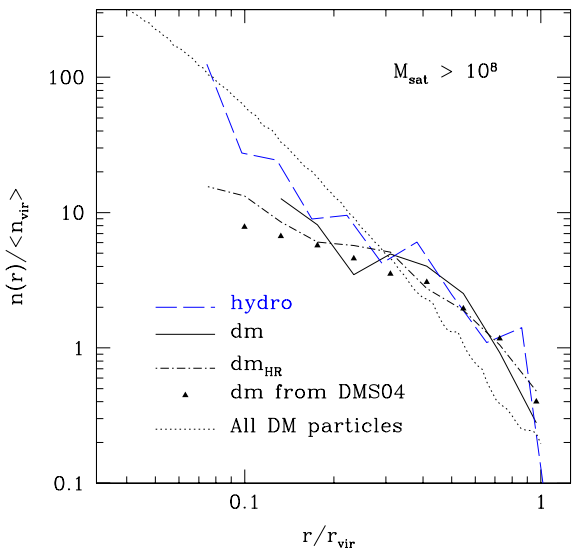}
\caption{Substructure radial density profiles for different simulations.}
\label{fig:profsub}
\end{figure}

Figure \ref{fig:profsub} shows the relative number density of surviving satellites
 $M>2 \times 10^8$ at z=0, in several pure dark matter simulations 
and our hydro simulation.
In all {\it dm} cases, the satellites are more extended than the overall
mass distribution (dominated by the smooth dark matter background). This
effect becomes larger inside of half the virial radius. 
Within this region, the dark matter only simulations reveal 
a much flatter number density distribution.
The satellite distribution in the hydro simulation is much
steeper and it is rather similar to 
the smooth mass distribution even if also in this case the overall radial
distribution of the subhalos is more extended than the underlying
dark matter distribution as shown clearly by the excess in the number density 
of subhalos at $0.5<r/r_{vir}<1.0$.

The inner core distribution found in the $dm$ run almost disappears and the
satellite profiles are well fitted by an NFW like function even if they are still
less concentrated than the overall mass distribution (in agreement with
findings from hydro simulations on cluster scales by Nagai \& Kravtsov
(2005)); the concentration parameter ($r_{vir}/r_c$) for the satellite
distribution is $\approx 6.5$, where the one for the smooth DM distribution in
the range $0.07<r/r_{vir}<1.0$ is $c_{DM}=9.6$.


\subsection{Resolution tests}

\begin{figure}
\centering
\epsfxsize=\hsize\epsffile{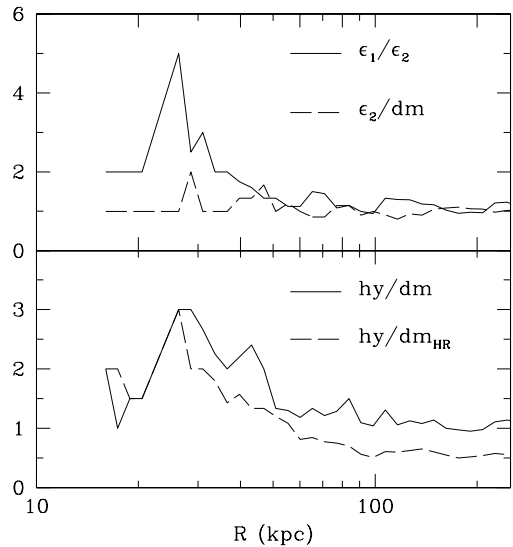}
\caption{ Effect of resolution on the number of subhaloes. {\it Upper panel}:
ratio between the number of DM subhaloes ($M> 2 \times 10^8 \Msun$) in the hydro run for two 
different values of the star particles softening: $\epsilon_1=0.2$ kpc and
$\epsilon_2 = 1.5$ kpc as function of distance from the center. The dashed line
is the same quantity but for the hydro run with $\epsilon = \epsilon_2$ and
the {\it dm} run. {\it Lower Panel}: the ratio between the number of DM subhaloes
($M> 10^7 \Msun$) in the hydro run and in the high resolution $dm_{HR}$ (dashed line) and
low resolution {\it dm} (solid line) runs.
}
\label{fig:resol}
\end{figure}

In the hydro simulation we have a higher mass resolution due to the presence of
baryons; the mass of the star and gas particles is roughly $20\%$ of
the mass of the DM particles, this means that in principle the lower number of
subhaloes close to the center that we found in the N-Body simulation can be
partially due to an overmerging phenomenom that depends on the
resolution. In the lower panel of Figure \ref{fig:resol} we compare the number of
subhaloes as a function of radius between the hydro
simulation and the highest resolution dark matter only run ($dm_{HR}$) which has a mass
resolution a factor 4 better than the low resolution case.
Figure \ref{fig:resol} demonstrates that the increase in the number of
substructure haloes in the hydro run is apparent even when compared to
the higher resolution N-Body simulation.

The second test that we have performed is related to the force resolution
(gravitational softening) adopted for the star particles. Stars are
concentrated in the inner few kpc of the galaxy and a 
small softening must be chosen to correctly follow
the dynamical evolution of such high density regions.
We have run the hydro simulation for two different values of the star gravitational
softening ($\epsilon$): $\epsilon_1 = 0.2$ kpc and $\epsilon_2 = 1.5$ kpc. 
Results are presented in the upper panel of Figure \ref{fig:resol} where we
show the difference in the number of subhaloes as a function of the distance
from the center. The effect of softening is important - too large a value leads
to tidal disruption of satellite galaxies since they have artificially shallow
central potentials and are hence less bound. 
The larger stellar softening ($\epsilon_2 = 1.5$ kpc) erases the stabilizing
effect of the stars inside the satellites and the radial distribution of
surviving subhaloes is identical
to the pure dark matter case (see the dashed line in the upper 
panel of Figure \ref{fig:resol}).

\section{Comparison with observations}
\label{sec:Obs}

For this comparison we used the data for the Local Group as compiled 
by Mateo et al. (1998). In this sample only
relatively massive satellites with an estimated rotational velocity or a
three-dimensional velocity dispersion of stars greater than 10 km sec$^{-1}$
are considered. In order to simplify the comparison even further, as
Klypin et al. (1999) we have considered the number of satellites within a
radius of 280 kpc from the Milky Way and Andromeda, 
which is close to the expected virial radius of these galaxies.

Even though the remaining gas particles are turned into collisionless tracers
at $z=1.5$, this does not affect the kinematical properties of the satellites
at z=0. All of the satellites that end up within the virial radius of the galaxy
halo have formed prior to this epoch. Subsequent loss of the remaining gas particles
to stars or by stripping processes would not lead to significant changes to the 
satellite structural parameters.

\begin{figure}
\centering
\epsfxsize=\hsize\epsffile{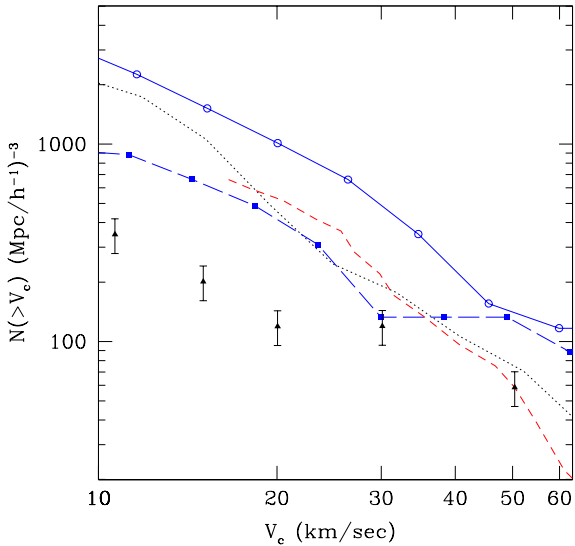}
\caption{Cumulative circular velocity distribution function of
satellites. Black solid line represents the results for the DM only
simulation, blue solid line is the VDF for stars (visible) haloes, the dashed
blue line is for DM haloes. Black triangles with error bars show average
results for Milky Way and Andromeda Satellites (Mateo 1998 \& Klypin et al
1999). The red dashed line are results for a \LCDM~  dissipationless galaxy
obtained by Klypin et al  (1999). 
}
\label{fig:Vcirc}
\end{figure}

In Figure \ref{fig:Vcirc} we show the cumulative velocity distribution
function (VDF) of satellites: the number of satellites per unit volume and per
central object with internal circular velocity larger than a given value
$V_{c}$. We show the VDF for different runs: the solid line represents
the results for the {\it dm} simulation, as already known 
the CDM models over-predict the number of satellites for
$V_{c} < 30$ km sec$^{-1}$. The dashed line represent the results from Klypin et
al 1999, that were originally compared with the data from Mateo et al 1998.
The simulation results agree quite well above 20 km sec$^{-1}$,
our higher resolution pure dark matter simulations
allow us to follow the distribution of subhaloes to lower circular velocities
($ V_{c}> 10$ km sec$^{-1}$).

The open circles connected by a solid line shows the VDF of dark matter haloes
for the hydro run.
The numbers of satellites in the range $15 < V_{c}< 30$ km sec$^{-1}$ is
increased, and now also for $V_{c} > 35$ km sec$^{-1}$ there is an over
abundance of satellites. 
We have also computed the VDF for satellites in the hydro run considering only 
their stars, shown by squares connected with the long dashed line.
This is a fairer comparison since we are comparing stellar kinematics in each case.
Stars are more concentrated than DM, so they trace the dynamics within a smaller
central region, this explains why the VDF for stars is below the DM
one over the whole velocity range (Hayashi et al 2004)
but this effect is much to small to reconcile the simulations with the
observations of Local Group satellites.

\begin{figure}
\centering
\epsfxsize=\hsize\epsffile{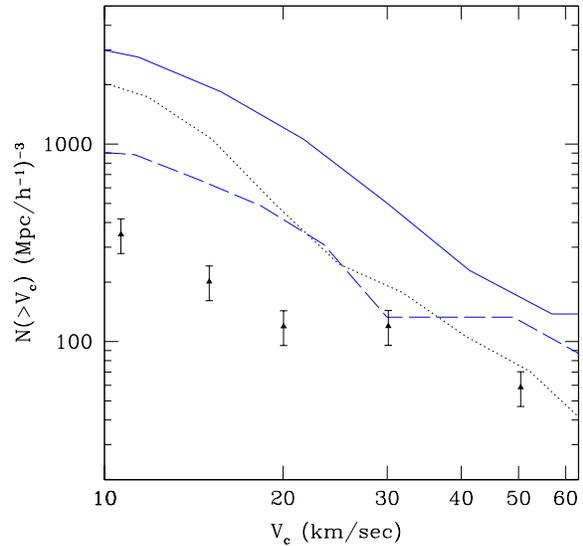}
\caption{A similar plot to figure \ref{fig:Vcirc}: the solid (blue) line is the
stellar  satellite VDF in absence of SN feed back, the dashed line is with feedback
  turned on, black dotted line represents the results for the DM only simulation. 
}
\label{fig:Vcirc.FB}
\end{figure}

Another clear difference between the two VDFs is that 
we have no stellar satellites with $V_{c}<15$ km sec$^{-1}$, which is due
to the supernovae feedback and UV background in the simulation. 
To confirm this we have run a full hydrodynamical simulation without 
these external feedback sources, shown in fig. \ref{fig:Vcirc.FB}, where the
VDF for the no feedback case continues to rise within $V_{c}<15$.
With the weak feedback used in this paper
hydrodynamic cosmological simulations produce VDF's which lie above the pure dark matter
results and the discrepancy to Local Group observations becomes
even larger. Runs with strong feedback from reionization are
able to produce realistic VDF's, we will present such runs
in a forthcoming paper (Macci\`o et al. 2005b).


\section{Lensing Analysis}
\label{sec:LensAn}

Quasars that are being gravitationally lensed into multiple images have
recently been used to place limits on the surface density of cold dark matter subhaloes
(Mao \& Schneider 1998, Metcalf \& Madau 2001, Chiba 2002,
Metcalf \& Zhao 2002, Dalal \& Kochanek 2002, Chen Kravtsov \& Keeton 2003,
Bradac et al. 2004, Mao et al 2004, Amara et al. 2004). 
Small mass clumps that happen to lie near the images affect the observed magnification
ratios. The question arises as to whether these observations are
compatible with distortions expected to occur from dark matter substructures
and satellite galaxies within the $\Lambda$CDM model. 
In the following we will first
present our lensing code, then we will briefly recall the main features of
the so called ``Cusp Relation'' and finally we will report our results.

\subsection{Lensing Simulations}
\label{sec:Lenscode}

Our ray shooting code is described in detail in Macci\`o 2005, here we will
only summarize its mean features.
The galaxy is centered in a cube of length 0.6 Mpc
and we study three lense images, obtained by projecting the particle
positions along the three coordinate axes. 
We then divide the projected density field $\Sigma$ by the critical 
surface mass density for lensing
\begin{equation}
\Sigma_{cr} = { c^2 \over {4 \pi G}} { D_S \over {D_L D_{LS}}} ~,
\end{equation}
thus obtaining the convergence $k$. Here $c$ is the speed of light, 
$G$ is the gravitational constant, while
$D_L$, $D_S$, $D_{LS}$ are the angular-diameter distances between lens and
observer, source and observer, lens and source, respectively.
In the following,
we adopt $z_L$ = 0.3 for the lens redshift and $z_s=3.0$ for the source redshift.

The deflection angle due to this 2D particle distribution, on a given point
$\vec x$ on the lens plane reads:
\begin{equation}
\vec \alpha(\vec x) = \sum_{j=1}^N {{4 G } \over 
{c^2}} { {m_j} \over { \vert \vec x - \vec y_j \vert}} ~.
\label{eq:alpha}
\end{equation}
Here $\vec y_j$ is the position of the $j$-th particles and $N$ is the total
number of particles.

Since a direct summation requires a long time, we speed up the code by using a
P$^3$M--like algorithm: the lens plane was divided into 256$\times$256 cells and
direct summation was applied to particles belonging to the same cell of $\vec
x$ and for its 8 neighbor cells. Particles in other cells were then seen as a
single particle in the cell baricenter, given the total mass of the particles inside the cell.
The deflection angle diverges when the distance between a light ray and a
particle is zero. To avoid this unwanted feature we introduce a softening parameter,
$\epsilon$ in equation (\ref{eq:alpha}); the value 
$\epsilon$ is tuned to the resolution of the current simulation.

We start to compute $\vec \alpha (\vec x)$ on a regular grid of 
4096$\times$4096 test rays that cover the central quarter of the lens plane
(multiple images form very close to the lens center). This resolution does not
provide enough pixels in the inner regions affected by strong lensing to
model lensing properties in the correct way. The resolution is increased 
by extracting the central region where strong lensing is occurring and using bilinear interpolation 
to calculate the relevant  quantities to higher resolution. Our final resolution 
is equivalent to a bundle of 16384x16384 light rays.

The relation between image and sources positions is given by the lens
equation:
\begin{equation}
\vec y = \vec x - \vec \alpha (\vec x) 
\label{eq:len}
\end{equation}
and the local properties 
of the lens mapping are then described by the Jacobian matrix
of the lens equation,
\begin{equation}
A_{hk}(\vec x ) = { {\partial y_h } \over {\partial x_k}} = \delta_{hk} - 
{ {\partial \alpha_h \over \partial x_k}}
\end{equation}
and the magnifications factor $\mu$ is given by the Jacobian determinant of $A$,
\begin{equation}
\mu(\vec x)  = {1 \over { \det A}} = [ A_{11}(\vec x)A_{22}(\vec x)- 
A_{12}(\vec x) A_{21}(\vec x)]^{-1} .
\end{equation}

Finally, the Jacobian also determines the location of the critical curves $\vec x_c$
on the lens plane, which are defined by $\det A(\vec x_c) = 0$. Because of the
finite grid resolution, we can only approximately locate them by looking for
pairs of adjacent cells with opposite signs of $\det A$. Then the lens 
equations
\begin{equation}
\vec y_c = \vec x_c - \vec \alpha(\vec x_c),
\label{eq:lens}
\end{equation}
yields the corresponding caustics $\vec y_c$, on the source plane.

To find the images of an extended source, all image-plane positions 
$\vec x$ are checked to see if the corresponding entry in the map table 
$\vec y$ lies within the source: i.e. for a circular source with radius
$r_c$ and centered in $(y^c_1;y^c_2)$ it is checked if:
\begin{equation}
 (y_1-y_1^c)^2 + (y_2-y_2^c)^2  \leq r_c^2,
\end{equation}
where $(y_1,y_2)$ are the components of the vector $\vec y$.
The sources are modeled as circles with a radius of $60$ pc according to
Amara et al (2004).
Those points fulfilling the previous equation are part of one of the source
images and are called image points. We then use a standard {\it friends-of-friends}
algorithm to group together image points within connected regions, since they
belong to the same image.

\begin{figure}
\centering
\epsfxsize=\hsize\epsffile{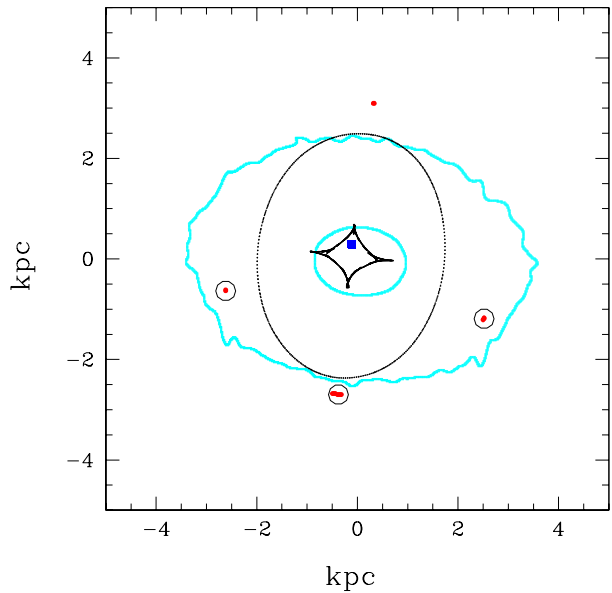}
\caption{Basic lens configuration. The caustic surface is shown as a black line and critical
 curves are shown as cyan lines. The four images that are usually observed
 are shown as red dots, and the three that are used to inspect the cusp
 relations are the ones inside the small black circles. The blue square is the
 source position. The softening adopted for this lens map is 0.5 Kpc. 
}
\label{fig:lens}
\end{figure}

A typical lens configuration is shown in Figure \ref{fig:lens} where critical
lines, caustic lines, source and images position are indicated.
The images within the black circles correspond to the those 
selected for the cusp relation investigation.

\subsection{Cusp Caustic Relation}
\label{sec:cusp}

There are basically three configurations of four-image  systems: fold, cusp,
and cross. In this paper we will mainly concentrate on the {\it cusp} 
configuration, that corresponds to a source located close to the cusp of the
inner caustic curve. The behavior of gravitational lens mapping near a cusp
was first studied by Blandford \& Narayan (1986), Schneider \& Weiss (1992) and
Mao (1992) and Zakharov (1995), 
who investigated the magnification properties of the cusp images
and concluded that the sum of the signed magnification factors of the three
merging images approaches zero as the source moves towards the cusp. In other
words (e.g. Zakharov 1995) :
\begin{equation}
R_{cusp} = {{ \mu_A + \mu_B + \mu_C} \over { \vert \mu_A \vert + \vert 
\mu_B \vert + \vert\mu_C \vert}} \rightarrow 0, ~~for ~~~~\mu_{tot} ~\rightarrow \infty
\label{eq:cusp1}
\end{equation}
where $\mu_{tot}$ is the unsigned sum of magnifications of all four images,
and A,B \& C are the triplet of images forming the smallest opening angle. By
opening angle, we mean the angle measured from the galaxy center and being
spanned by two images of equal parity. The third images lies inside such an angle.
This relation is an asymptotic relation and holds when the source approaches
the cusp from inside the inner astroid caustic.

Since we know the lens position and the source position the procedure of
finding the cusp images is straightforward, we have identified the triplet of
images belonging to the smallest opening angle, we have seen that the cusp
images are better identified using as opening angle measured from 
the fourth image and being spanned by two images of equal parity. We have
used this method only to find the cusp images, instead for testing the cusp relation 
(see eq. \ref{eq:cusp2}) we have used the opening angle as defined from
the center of the galaxy ($\Delta\theta$).

\begin{figure}
\centering
\epsfxsize=\hsize\epsffile{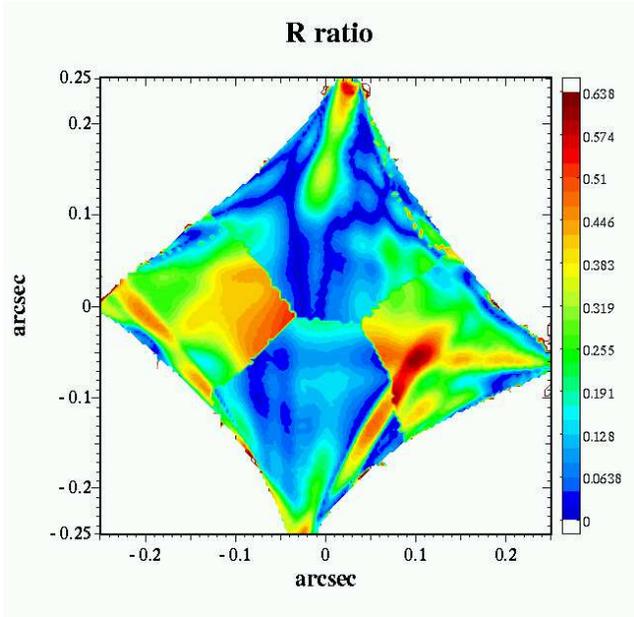} 
\caption{Value of the quantity $R_{cusp}$ in the four images region of the
  source plane, here a softening of $0.3$ Kpc is used.}
\label{fig:Rcusp}
\end{figure}

Approximately 25000 lens systems are generated with the source position inside
the astroid caustic. Figure \ref{fig:Rcusp} indicates in color the value of $R_{cusp}$
for all the different source positions; the apparent discontinuities originate
from different image identifications. In the very center of the caustic the
cusp relation is not well defined (what you have is mostly a ``cross
relation'', four images situated at the vertices of a cross centered on the
cusp center), as the source moves in the direction of the minor or major axes
we choose different subsets of three cusp images and therefore the
discontinuity arises.

\section{Effects of substructures}
\label{sec:LensRes}

Because of the finite size (discreteness) 
of the particles in the simulations there is a
significant amount of shot noise in the surface density estimate, which can
affects the lensing properties. The usual approach (Bradac et al. 2004, Amara
et al. 2004) is to use a Gaussian kernel to smooth the surface density, in
Amara et al. (2004) a detailed study of the impact and calibration of the
smoothing was presented, their main conclusion was that a smoothing of 0.5
kpc is suitable for studying the cusp relation properties without losing any
spatial information.

In order to try to disentangle the effects of bound substructures to the spurious
effect produced by single finite particles, we have adopted a novel approach. 
We have have tried to
{\it remove} substructures from the lensing halo to see how this can change the cusp
relation. Our removal procedure works as follows: first we identified all bound
substructure using SKID (see section \ref{sec:sub}), then each particle
belonging to any of the subhaloes is rotated around the center of the
galaxy using three random Euler angles. In order to avoid thin circular shells
of particles we added a random $\pm 10 \%$ error to each distance.
We want to emphasize that we do not physically remove any substructures,
because this will change the overall properties of the lens: mass, 
density profile, etc.

\begin{figure}
\centering
\epsfxsize=\hsize\epsffile{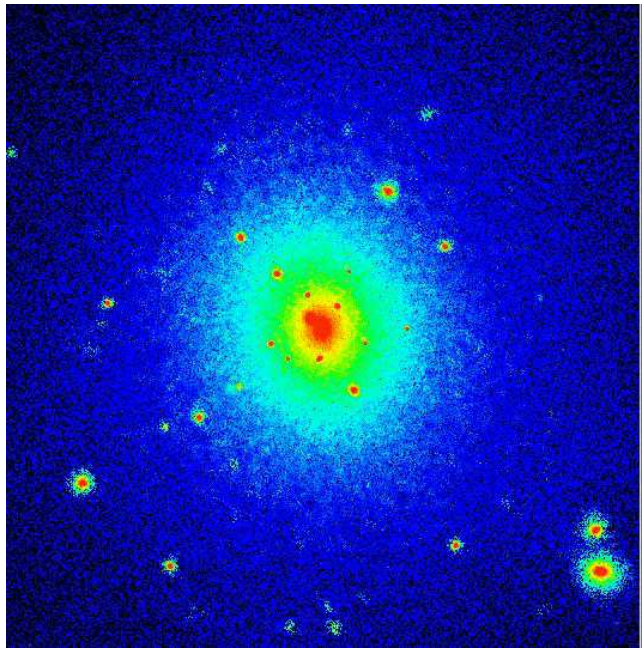}
\caption{Density map of the mass distribution within the full hydrodynamical simulation. 
The size of the box is $200$ kpc.
}
\label{fig:den1}
\end{figure}

\begin{figure}
\centering
\epsfxsize=\hsize\epsffile{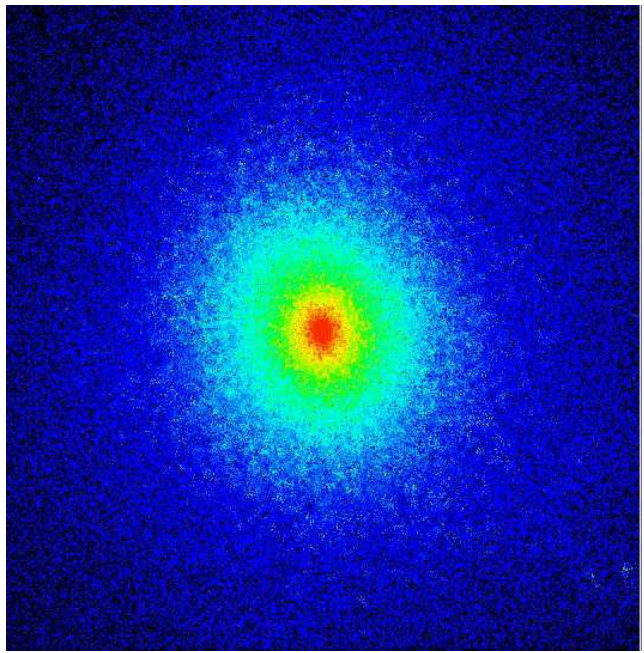}
\caption{
Density map of the smoothed mass distribution after randomizing the positions of
the particles within the substructure haloes.}
\label{fig:den2}
\end{figure}

\begin{figure}
\centering
\epsfxsize=\hsize\epsffile{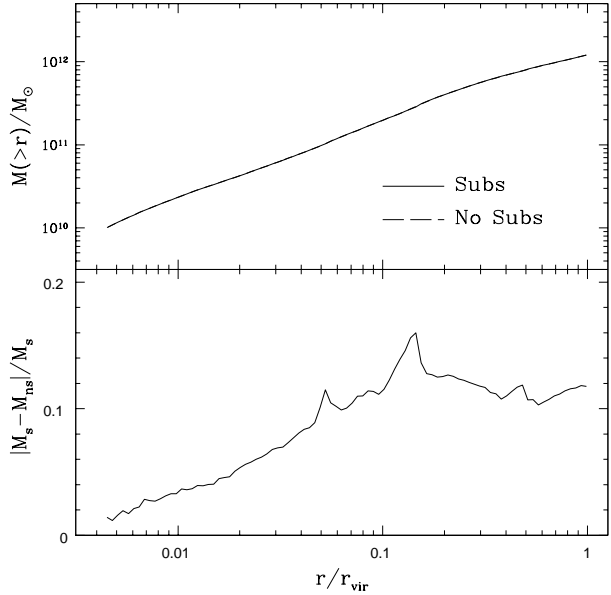}
\caption{Mass versus radius before and after the removal of substructures, the
  overlap between the two curves is very good, in the lower panel the
  residuals are shown}.
\label{fig:removal}
\end{figure}

This procedure allows us to smooth out only the substructures leaving unaltered
all the main features of the primary lens. In Figure \ref{fig:removal} (upper
panel) we show the integral radial mass profile before and after the {\it removal} 
of the substructures, not only the total mass is preserved but also its radial
profile, this means that all the lensing properties of the galaxy remain
unchanged (critical curves, caustics, position of images) allowing
us to make a one to one comparison between the two lensing maps.
In Figure \ref{fig:den1} and Figure \ref{fig:den2} the density map of the
galaxy is shown, with and without substructures respectively. 

The cusp relation defined by equation \ref{eq:cusp1} holds when the source is close
to the cusp. As soon as the source moves away from the cusp
deviations from $R_{cusp}=0$ are observed, even for the smooth lens model.
On the other hand the closer the source is to the cusp, the smaller is the angle spanned from
the three images, so in order to take into account also the position of the
source in evaluating the cusp relation it is better to define $R^0_{cusp}$
as (see also Amara et al. 2004):
\begin{equation}
R^0_{cusp} = {{2 \pi} \over  {\Delta \theta} } R_{cusp}.
\label{eq:cusp2}
\end{equation}
where $\Delta\theta$ is the opening angle spanned by the two images with
positive parity defined from the center of the galaxy.

With this new definition of $R_{cusp}^0$ a set of three images is said to
violate the cusp relation if $R_{cusp}^0 > 1$.
This makes the comparison between simulations and observations much more straightforward.

\begin{figure}
\centering
\epsfxsize=\hsize\epsffile{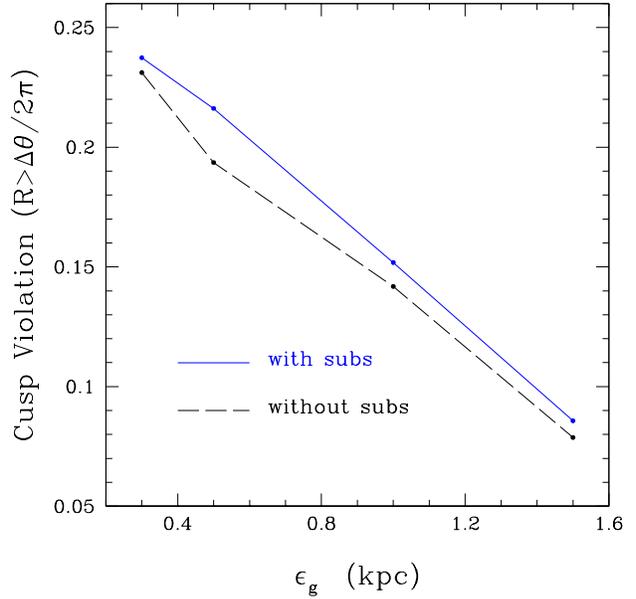}
\caption{Fraction of total number of sources that violate the cusp relation as
  a function of the
  Gaussian smoothing length $\epsilon_g$. Solid line is for the whole galaxy, dashed line
  for the galaxy {\it without} substructures (see text for its definition).
}
\label{fig:Cusp2}
\end{figure}

The differences in the reduced cusp relation violation in the two cases are shown in
Figure \ref{fig:Cusp2}, where we plot the number of sources that violate the 
{\it reduced} cusp relation as a function of the Gaussian smoothing scale ($\epsilon_g$).
For both the lens models, the number of sources that violate equation \ref{eq:cusp2}
decreases with the smoothing length because the effect of smoothing is
to both reduce the impact of substructures and the noise introduced by single particles.

The difference between the two is not so large, with a maximum for
$\epsilon_G = 0.5$ kpc, where the number of violations grows from
19\% to 23\%,
this is because this value of $\epsilon_G$ is large enough to
cancel the Shot noise, but not large enough to smear out the subhaloes in the
simulation, in good agreement with the results of Amara et al.
For smaller value of $\epsilon_G$ the signal is almost completely dominated by Shot noise,
and for larger values we smooth too much, losing spatial information on
the surface density of the lens.

Figure \ref{fig:Cusp2} clearly shows that the impact of substructure in a mass
range $10^7-10^9$ is very weak in disturbing the cusp relation. 
The resolution achievable with current numerical simulations is still too poor
to extend the analysis to a lower mass range of subhaloes, analytic arguments
or semi-analytic prescriptions must be used (Macci\`o \& Miranda 2005).

\begin{figure}
\centering
\epsfxsize=\hsize\epsffile{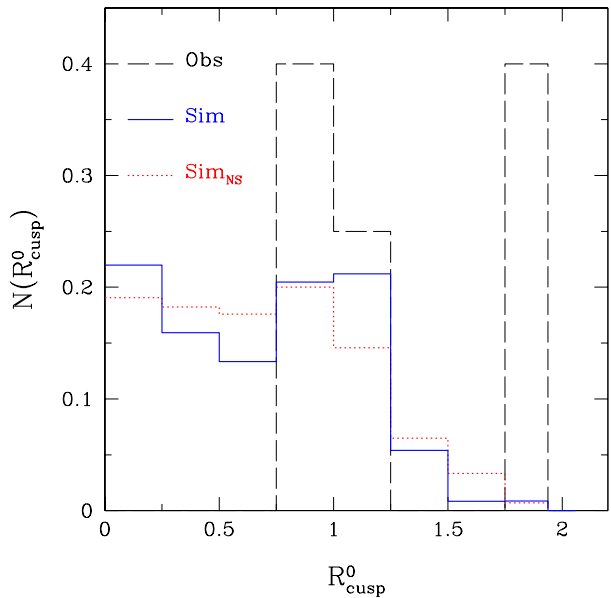}
\caption{Distribution of $R^0_{cusp}$ values. The solid and dotted lines show
  the simulation results before and after the substructure removal. The long
  dashed line represents the observational data.
}
\label{fig:histo}
\end{figure}

Nevertheless a tentative comparison with observation can be made; 
there are 5 observed cusp caustic lenses systems: 
{\noindent B0712+472 (Jackson et al
1998), B2045+265 (Koopmans et al 2003), B1422+231 (Patnik \& Narasimha 2001),
RXJ1131-1231 (Sluse et al. 2003) and RXJ0911+0551 (Keeton et al 2003); the
first three are observed in the radio band, the last two in optical and IR.
Three of them violate the reduced cusp relation (i.e. $R_{cusp}> \Delta \theta
/ 2 \pi$). This means a 60\% violation, that is significantly larger than the
15-25\% we found for our simulations.} The sources size used in this work
($60 \rm pc$) allows us to make a direct comparison mainly with QSO observed in
radio than in optical or IR, even in this case we have a violation of the reduced
cusp relation in 2 objects over 3, that means 66\% violation.
Figure \ref{fig:histo} shows the distribution of the values of the reduce
cusp relation $R_{cusp}^0$ both for data and simulates systems (with $\epsilon_G=0.5
$ kpc). 
Simulation results are unable to reproduce the high value tail that arises 
in the observational data; again it is possible to
see that the effect of subhaloes is very weak in disturbing the cusp
relation, and they only marginally enhance the number of systems with $R_{cusp}^0>1$.

As already discussed above, the difference between data and simulation results
can arise from the current resolution limitations in the Numerical simulations, 
we expect to have many more small DM haloes close to the
center of the galaxy with masses in the range $10^{-3}-10^6 ~ \Msun$, these haloes will be
very concentrated and so they can more easily survive the tidal force of the
central halo and if one of these haloes is close enough to one of the images it
can perturb its magnification and so violate the cusp relation.
(the smaller is the mass of the subhalo the closer it must be to the 
projected image position).

Another possible explanation for the observed cusp relation violation, can be
ascribed to the effect of the all subhaloes that are in the galactic space
along the line of sight of the lens (Metcalf 2004); the importance of
these intergalactic haloes with mass $< 10^8 ~ \Msun$ depends on the
radial profile of the dark matter haloes and the primordial power spectrum at
small scales.

What is clear from our analysis is that the cusp relation violation can not be
due to substructures in the primary lens with a masses above $10^7\Msun$. This
is true even though the inclusion of baryons has increased the projected numbers 
of subhaloes by a large factor.


\section{DISCUSSION AND CONCLUSIONS}
\label{sec:Concl}

Using a high resolution hydrodynamic galaxy formation simulation 
we have studied the number
density and the spatial distribution of subhaloes within a Milky Way 
sized CDM halo.
Baryons can concentrate at the halo centre via two mechanisms: The shock
heated gas can radiate away energy causing the particles to fall inwards. Cold
gas clouds can accrete along filaments, colliding and coalescing
with existing gas at the halo centre. The slowly infalling baryons has
the effect of increasing the density of the dark matter via adiabatic
contraction. The overall effect of including baryons is to steepen the mass
profile to nearly isothermal with $\rho(r)\propto r^{-2}$ from a shallower
$r^{-1}$ cusp.

As already predicted by several (semi)analytic studies, our simulations show
that this central concentration of baryons enables subhaloes to better withstand
the tidal forces generated by the main halo. This leads to an increase
in the total number of subhaloes, especially within the inner third
of the virial radius, such that they follow the overall mass distribution.
This is in excellent agreement with the distribution of galaxies within
galaxy clusters. However on the scale of Galactic mass haloes, this 
increases the discrepancy between the numbers of surviving satellites in CDM 
models and the flat luminosity function observed within the Local Group.
A possible solution to this problem may come from the early reionization of
the universe from early structure formation (Madau \& Rees 2001).
Reionization raises the entropy of the gas that is required to fuel galaxy
formation, preventing it from accreting into small dark matter haloes and lengthening the
cooling time of that gas which is accreted (Bullock et al 2001, Somerville
et al 2002, Ricotti, Gnedin \& Shull 2002).
Moreover as suggested by Benson \& Madau (2003) winds from pre-galactic starbursts and
mini-quasars may pollute the intergalactic medium (IGM) with metals and raise
its temperature to a much higher level than expected from photoionization and
so inhibit the formation of early galaxies.

A detailed study of the impact of reionization on the formation of dwarf
galaxies in hydrodynamic simulation will be presented in a forthcoming paper
(Macci\`o et al 2005b), where we show that with an appropriate choice of the
reionization parameters and modeling it is possible to reconcile
observational data with the steep mass function of haloes and subhaloes
within the cold dark matter model.

In the second part of this work we have explored the consequences of the increased
numbers of satellites for multiply lensed quasar images by foreground
galactic mass haloes.
In particular, the signatures on the violation of the so called cusp relation.
Previous work have reached different conclusions on this issue: Bradac et
al (2004), found an agreement between simulations and observations but their
results were limited by the low resolution of their simulations. 
Amara et al (2004) (see also Mao et al 2004) have shown that is hard to reconcile the observed
high number of cusp relation violation with results from simulations
moreover they have also show that Shot noise due to the discreteness of the
simulation (where every particle is in principle a substructure) plays an
important role in changing the properties of the lens map.
In order to disentangle the effect of substructures from the effect of having
an intrinsic discrete distribution of matter we compared results between
haloes with and without any substructure present.
We have developed a new technique to remove the satellites from the simulation
with out changing the overall matter distribution of the primary lens.
This analysis demonstrates that the impact on lensing of subhaloes 
in the mass range $10^7 - 10^{10} ~\Msun$ is very small. 
Also having a number of subhaloes which is about 8 times higher of the observed
one in this mass range, the number of multiple lensed QSO that show a
violation of the cusp relation (defined as eq. \ref{eq:cusp2}) is less than 24\%,
in contrast with an observed one of about 60\%.
Our results extend down to subhaloes masses of $10^7 \Msun$ due to the resolution limit
of our simulations. Even considering the impact of smaller masses for subhaloes
using an analytic approach does not help in solving the cusp relation problem
(Macci\`o \& Miranda 2005).
A possible  explanation could be that more variables must be taken into account in the lensing
analysis, such as all the (sub)haloes that lie along the line of sight between us
and the lens (see Chen et al. 2003, Metcalf 2004).

\section*{Acknowledgments}

The authors acknowledge J. Wadsley for development of the GASOLINE code and 
thank him for its use in this work and thank A. Kravtsov, D. Nagai, C. Mastropietro,
L. Mayer, M. Miranda and P. Saha for useful
discussions during the preparation of this work. We also thank the referee
Hongsheng Zhao for his comments. AM also thanks S. Kazantzidis
for the use of the {\it extract} software. JD is supported by the Swiss
National Science Foundation. All the numerical simulations were
performed on the zBox supercomputer 
(http://www-theorie.physik.unizh.ch/$\sim$stadel/) at the University of
Z\"urich.

\end{document}